\documentclass[aps,twocolumn,superscriptaddress,showpacs,floatfix]{revtex4-1}
\usepackage{amsmath,amssymb,graphicx}

\usepackage[utf8]{inputenc}
\usepackage[T1]{fontenc}
\usepackage{xcolor}

\IfFileExists{newtxtext.sty}
   {\usepackage{newtxtext,newtxmath}}
   {\IfFileExists{stix.sty}
      {\usepackage{stix}}
      {\IfFileExists{mathptmx.sty}
      {\usepackage{mathptmx}}{} } }

\usepackage{textcomp}

\usepackage{bm}

\IfFileExists{siunitx.sty}{\usepackage{booktabs,siunitx}}{}

\pdfoutput=1
\usepackage{color}
\definecolor{LinkColor}{rgb}{0.256,0.439,0.588}
\usepackage{hyperref}
\hypersetup{
   pdfauthor={Xiao Yan Xu, Stefan Wessel, and Zi Yang Meng},
   pdftitle={Competing pairing  channels in the doped honeycomb lattice Hubbard model},
   colorlinks=true,
   citecolor=LinkColor,
   linkcolor=LinkColor,
   urlcolor=LinkColor
}

\renewcommand{\vec}[1]{\mathbf{#1}}

\usepackage{pifont}

\begin{document}

\title{Competing pairing  channels in the doped honeycomb lattice Hubbard model}

\author{Xiao Yan Xu}
\affiliation{Beijing National Laboratory for Condensed Matter Physics and Institute
of Physics, Chinese Academy of Sciences, Beijing 100190, China}
\author{Stefan Wessel}
\affiliation{Institute for Theoretical Solid State Physics, JARA-FIT and JARA-HPC,
RWTH Aachen University, 52056 Aachen, Germany}
\author{Zi Yang Meng}
\affiliation{Beijing National Laboratory for Condensed Matter Physics and Institute
of Physics, Chinese Academy of Sciences, Beijing 100190, China}

\begin{abstract}
Proposals for superconductivity emerging from correlated electrons in the doped Hubbard model on the honeycomb lattice range from chiral $d+id$ singlet  to  $p+ip$ triplet pairing,
depending  on the considered range of doping and interaction strength, as well as the approach used to  analyze the pairing instabilities.
Here, we consider  these scenarios  using large-scale  dynamic cluster approximation (DCA) calculations to  examine the evolution in the leading pairing symmetry
from weak to intermediate coupling strength.
These calculations focus on doping levels around the van Hove singularity (VHS) and are performed using DCA simulations with an  interaction-expansion continuous-time quantum Monte Carlo cluster solver.
We calculated explicitly the temperature dependence of different uniform superconducting pairing susceptibilities and found a consistent picture emerging upon gradually increasing the cluster size: while at weak coupling the $d+id$ singlet pairing  dominates close to the VHS filling, an enhanced tendency towards $p$-wave triplet pairing upon further increasing the interaction strength is observed. The relevance of these systematic results for existing proposals and ongoing pursuits of odd-parity topological superconductivity are also discussed.
\end{abstract}


\date{September 2, 2016}
\maketitle

\section{Introduction}
Many aspects  of the fascinating physics of the low-energy Dirac electrons in graphene can be explored based on
noninteracting tight-binding models on the honeycomb lattice, in particular close to  charge neutrality,
where the effects of the electronic interactions are delayed to the strong-coupling regime due to  a vanishing density
of states (DOS) at low energies.   At finite doping, however,
the presence of even weak interactions among the electrons is predicted by several studies to lead to  new collective behavior. Of particular recent interest are interaction-driven instabilities towards  unconventional superconductivity in doped honeycomb systems~\cite{Uchoa_2007_146801,Honerkamp_2008_146404, Ma_2011_,Black-Schaffer_2007_,Nandkishore_2012_158, Kiesel_2012, Wang_2012_,Wu_2013_,Black-Schaffer_2014_,Black-Schaffer_2014_423201,Faye_2015_085121,Roy_2010, Faye_2016,Gu_2013_,Gu2014,Roy2014,Kunst2015}.
Some early studies concluded that superconductivity might not be stable with respect to charge or spin order  for the
basic Hubbard model on the honeycomb lattice~\cite{Honerkamp_2008_146404, Ma_2011_},  and a possible  quantum liquid
state has been suggested recently for the van Hove singularity (VHS) filling~\cite{Jiang2014}.
In most of the recent theoretical studies, however,
a general tendency towards some flavor of superconductivity upon doping the honeycomb lattice is indeed observed.
However,  various proposals on the nature of the emerging superconducting state and the stability range of competing pairing channels still lead to a mosaic of different scenarios.
Several mean field theory and renormalization-group (RG) calculations predict chiral $d+id$  singlet superconductivity
to emerge in the weak-coupling region upon doping towards or onto the VHS, which corresponds to electronic densities of
$n=3/4$ and $5/4$ for the Hubbard or related models with explicit spin exchange terms or extended interactions~\cite{Black-Schaffer_2007_,Nandkishore_2012_158, Kiesel_2012, Wang_2012_,Wu_2013_,Black-Schaffer_2014_,Black-Schaffer_2014_423201}.
Variational Monte Carlo simulations~\cite{Pathak_2010_} also showed a chiral $d$-wave solution over a wide range of doping.
The $d$-wave pairing state in this scenario is related to enhanced antiferromagnetic fluctuations near half filling as well as the VHS-increased DOS.

On the other hand, a recent study using the variational cluster approximation (VCA) and cellular dynamical mean field theory (CDMFT) performed for larger values of the local repulsion,  found a stable $p$-wave triplet pairing state for a weak nearest-neighbor repulsion~\cite{Faye_2015_085121}, with possibly a coexisting Kekul\'{e} pattern~\cite{Roy_2010,Kunst2015,Faye_2016}.  A possible $p+ip$ pairing state was also reported at low filling from determinantal quantum Monte Carlo studies; however, the sign problem poses restrictions on the accessible system sizes, interaction strengths and temperature ranges. In addition, Grassmann tensor renormalization calculations have  been performed~\cite{Gu_2013_,Gu2014}, and in Ref.~\cite{Gu_2013_}, a $d+id$ state is reported for the $t-J$ model, while for infinite local repulsion a $p+ip$ superconducting state, coexisting with ferromagnetic order, has been proposed for the Hubbard model at low doping~\cite{Gu2014}.
Hence, despite active pursuits, such deviations among the various proposals and employed methods show that a consistent
picture of possible superconductivity  even in the basic Hubbard model on the honeycomb lattice is still lacking,
apparently due to  competition among several possible low-energy states upon varying the doping or  interaction
strength. It  thus appears promising and necessary to examine this problem from the perspective of a method that allows
us to tune these parameters over a wide range while accounting for the growing local electronic correlations beyond the weak-coupling regime.

Here, we employ such an approach by providing results from large-scale dynamic cluster approximation (DCA)~\cite{Maier2005} calculations, with a focus on
 pairing susceptibilities to probe for uniform superconducting instabilities.
Upon systematically increasing the cluster size, we find that a consistent picture starts to emerge for the leading
pairing channels on the  honeycomb lattice Hubbard model from small- to medium-sized local interactions: while at weak coupling, chiral $d+id$ singlet pairing dominates close to VHS  filling, when the interaction becomes stronger, a tendency towards $p$-wave triplet pairing develops. Our calculations are performed with an  interaction-expansion continuous-time quantum Monte Carlo (CT-INT)
cluster solver~\cite{Chen2012,Chen2013a,Chen2013b,Meng2015}, keeping up to $24$ cluster sites (see Appendix~\ref{app:a} for details on the  CT-INT approach). Before discussing our results, we   provide details about the considered model and the DCA computational framework for calculating the pairing susceptibilities.

\section{Model and method}
The Hubbard model on honeycomb lattice has the Hamiltonian
\begin{eqnarray}
\hat{H} &=& \hat{H}_0 + \hat{H}_{I}, \nonumber\\
\hat{H}_0 &=& -t\sum_{\langle i,j\rangle\sigma}\hat{c}_{i\sigma}^{\dagger}\hat{c}_{j\sigma}-\mu\sum_{i\sigma}\hat{n}_{i\sigma}, \nonumber\\
\hat{H}_{I}&=&U\sum_{i}(\hat{n}_{i\uparrow}-\frac{1}{2})(\hat{n}_{i\downarrow}-\frac{1}{2}),
\label{eq:hamiltonian}
\end{eqnarray}
where $t$ denotes the hopping amplitude between nearest neighbor sites $\langle i,j\rangle$, $\mu$ is the chemical potential that controls the electronic density,  and $\hat{n}_{i\sigma}=\hat{c}_{i\sigma}^{\dagger}\hat{c}_{i\sigma}$ is the number operator for spin flavor $\sigma$ on the $i$-th lattice sites. Furthermore, $U$ denotes the onsite Coulomb repulsion. Longer-ranged interaction will not be considered here, and  at finite doping, especially close to the VHS, screening plays an important role and cuts off the long-ranged tail of the Coulomb potential~\cite{Black-Schaffer_2014_423201}.

\begin{figure}
\includegraphics[width=\columnwidth]{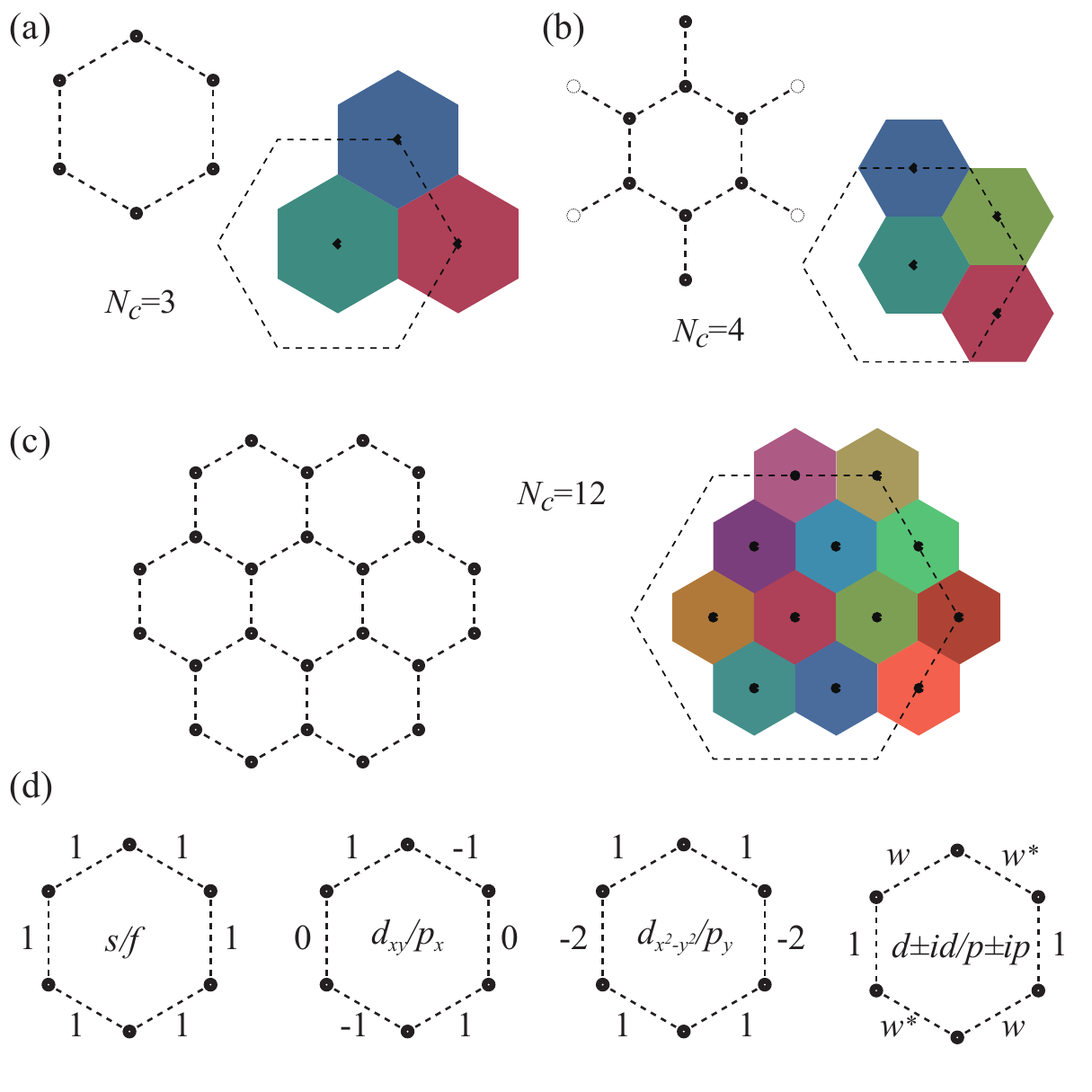}
\caption{(Color online) Real space clusters with (a) $N_c=3$, (b) $4$, and (c) $12$ unit cells along with  their
corresponding momentum patches within the Brillouin zone. The clusters $N_c=3$ and $12$ already retain the lattice
$D_{6h}$ symmetry, while  for the $N_c=4$  cluster we enforce this symmetry, as shown in (b), where the four outer
hollow sites and the two outer black sites are equivalent due to this symmetry. (d) The phase factors for difference nearest-neighbor pairing channels, where $s$, $d_{xy}$, $d_{x^2-y^2}$ and $d \pm id$ correspond to  singlet pairing states, while $p_x$, $p_y$, $p \pm ip$ and $f$ are triplet states. $w=\exp(\pm i2\pi/3)$ here .}
\label{fig:clusters}
\end{figure}

The DCA maps this original lattice model onto a periodic cluster, embedded into a self-consistently determined bath.
Spatial correlations within the cluster are treated explicitly, while those at longer length scales are described at the
dynamical mean-field level~\cite{Maier2005}. For this work, we have systematically employed three cluster sizes, shown
in Fig.~\ref{fig:clusters}, with $N_c=3,4$ and $12$ unit cells. The lattice $D_{6h}$ symmetry is enforced for the
$N_c=4$ cluster (see the figure caption), while the $N_c=3$ and $12$ clusters  explicitly retain this symmetry. For the largest cluster, we are able to study inverse temperatures up to $\beta t =40$ at a coupling of $U=2t$.
Compared to the widely studied square~\cite{Maier2006a,Maier2006b,Vidhyadhiraja09,YangSX11,Chen2011,Chen2012, Chen2013b,Gull2013} or triangular lattices~\cite{Lee2008,Chen2013a,Dang2015}, the DCA formalism needs to be
modified
for the honeycomb lattice, which is a  bipartite lattice with two sites per  unit cell. In particular, the single particle Green's function and the self-energy for spin flavour $\sigma$ are $2\times2$ matrices $G^{\sigma}_{\alpha,\beta}(\mathbf{K},i\omega_n)$ and $\Sigma^{\sigma}_{\alpha,\beta}(\mathbf{K},i\omega_n)$, with the band or orbital indices $\alpha,\beta=1,2$. We developed a  generic scheme for performing DCA calculations on such more complex lattices, 
with details provided in Appendixes~\ref{app:b} and \ref{app:c}.

\section{Pairing susceptibilities}
In order to probe for superconductivity with respect to different pairing channels, we consider appropriate pairing order parameters in real space,
\begin{equation}
\Delta_{\eta}(i)=\sum_{l}f_{\eta}(\boldsymbol{\delta}_{l})\left(\hat{c}_{i\uparrow}\hat{c}_{i+\delta_{l}\downarrow}\pm\hat{c}_{i\downarrow}\hat{c}_{i+\delta_{l}\uparrow}\right),
\end{equation}
where $\eta$ denotes the different pairing channels: $s$, $p$, $d$, $f$, $p\pm ip$ and $d\pm id$. Here,
$f_{\eta}(\boldsymbol{\delta}_{l})$ are form factors that correspond  to the pairing symmetry $\eta$ and are provided
explicitly in Fig.~\ref{fig:clusters}; $\boldsymbol{\delta}_{l}$ indicates  the pairing bonds (we restrict ourselves to
nearest-neighbor pairings, $l=1,2,3$), and $+$ and $-$ denote triplet and singlet states, respectively. The possible  pairing channels can be classified according to the irreducible representations of the  $D_{6h}$ point group of the  honeycomb lattice~\cite{Black-Schaffer_2007_,Black-Schaffer_2014_,Black-Schaffer_2014_423201,Faye_2015_085121}.
The corresponding uniform susceptibility for a pairing channel $\eta$ is then obtained in the imaginary-time formulation as
\begin{equation}
\chi^{\eta}(T)=\frac{1}{N}\int_{0}^{\beta}d\tau\sum_{ij}\left\langle \text{T}_{\tau} \Delta_{\eta}^{\dagger}(i,\tau)\Delta_{\eta}(j,0)\right\rangle.
\end{equation}
Transforming to momentum and frequency space and  normalizing by the form factors, we obtain
\begin{equation}
\chi^{\eta}(T) = \frac{1}{\beta}\frac{\sum_{\mathbf{p},\mathbf{p}',\mathbf{q}=0}\langle\Phi_{\eta}(\vec{k})|\chi(\mathbf{p},\mathbf{p'},\mathbf{q})|\Phi_{\eta}(\vec{k}')\rangle}{\sum_{\vec{k}}\langle\Phi_{\eta}(\vec{k})|\Phi_{\eta}(\vec{k})\rangle}
\end{equation}
which contains the form factor, written in vector form as
\begin{equation}
|\Phi_{\eta}(\vec{k})\rangle=\big(\sum_{l}f_{\eta}(\boldsymbol{\delta}_{l})e^{i\vec{k}\cdot \boldsymbol{\delta}_{l}},\ \mp\sum_{l}f_{\eta}(\boldsymbol{\delta}_{l})e^{-i\vec{k}\cdot\boldsymbol{\delta}_{l}}\big)^{\dagger}.
\end{equation}
Here, $\mathbf{p}=(\vec{k},i\omega_{n})$, $\mathbf{p'}=(\vec{k}',i\omega_{n}')$ and $\mathbf{q}=(\vec{q},i\nu_{m})$ are four-momenta containing both momentum and frequency.
In the following, we restrict ourselves to  uniform pairing states, corresponding to $\mathbf{q}=0$ and $\nu=0$; that
is,  we focus here on the pairing channel with respect to only the point group symmetry.
Directly comparing pairing susceptibilities $\chi^{\eta}(T)$ of different  channels is not a practical way to
numerically identify the leading pairing channel because, usually, the non-interacting pairing susceptibility masks the
interaction effects in the pairing susceptibility within the temperature range where the CT-INT simulation can be
performed. Two routes can be taken to overcome this issue. One possibility is to perform an eigenvalue analysis of the
pairing vertex secular equation~\cite{Maier2006a,Chen2013b,Meng2014,Meng2015}, which usually requires high quality  data
for the irreducible vertices and very low temperatures, such that the momentum dependence of the leading eigenvector
    does  indeed reflect the pairing symmetry of the superconducting ground state. In the other approach, one subtracts
    the decoupled part of the pairing susceptibility $\chi^{0}$ (the particle-particle bubble) from the interacting one,
    such that the effective pairing susceptibility $\chi_{\text{eff}}=\chi-\chi^{0}$, stemming  from the electronic correlations, can be extracted~\cite{Ma_2011_,Chen_2015_116402}.
Here, we adopt the latter scheme and thus extract the effective pairing susceptibilites 
\begin{equation}
\chi_{\text{eff}}^{\eta}(T) = \frac{1}{\beta}\frac{\sum_{\mathbf{p},\mathbf{p'},\mathbf{q}=0}\langle\Phi_{\eta}(\vec{k})|\chi(\mathbf{p},\mathbf{p'},\mathbf{q})-\chi^{0}(\mathbf{p},\mathbf{p'},\mathbf{q})|\Phi_{\eta}(\vec{k}')\rangle}{\sum_{\vec{k}}\langle\Phi_{\eta}(\vec{k})|\Phi_{\eta}(\vec{k})\rangle}.
\end{equation}
Further details on the calculation of the vertex function and the effective pairing susceptibilities within the DCA framework are provided in Appendixes~\ref{app:d} and \ref{app:e}.
The later also discusses the relation to the eigenvalue analysis of the pairing vertex. 

\begin{figure}[tp!]
\includegraphics[width=\columnwidth]{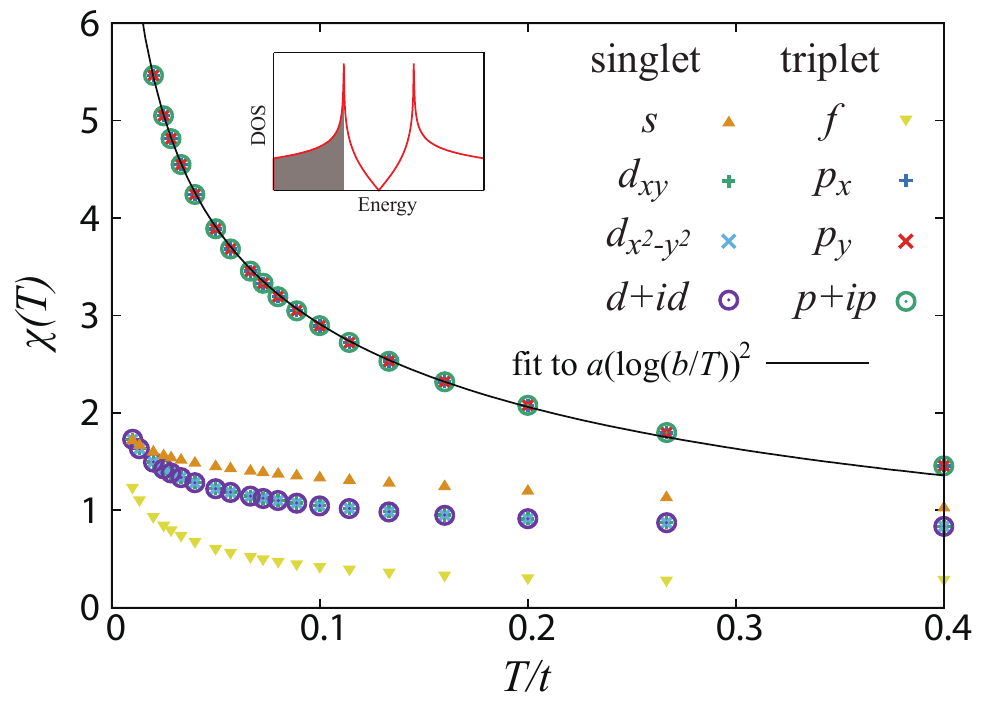}
\caption{(Color online) Non-interacting ($U=0$) pairing susceptibilities for various channels at the VHS filling. All channels will diverge at low temperature due to the logarithmic divergence of the DOS at the VHS. The $p$-wave susceptibilities exhibit the strongest increases, as indicated by the fit line. The inset  shows the peak in the DOS at the VHS for the density $n=3/4$ (gray).}
\label{fig:pairingu0}
\end{figure}

\section{Results}
Before discussing interaction effects, it is useful to examine
the bare $(U=0)$ pairing susceptibilities, which are shown in Fig.~\ref{fig:pairingu0} at the VHS density $n=3/4$. Due to the VHS in the bare DOS,  the bare $\chi(T)$ diverge logarithmically as $T\rightarrow 0$, with the strongest
divergence exhibited by the 
$p$-waves. This provides an important background to the pairing susceptibilities in the interacting case. Hence,  at finite $U$, the decoupled part of the pairing susceptibility 
$\chi^{0}(T)$  is subtracted in order to make the effective pairing susceptibilities $\chi_{\text{eff}}(T)$ manifest, as  mentioned above. 

\begin{figure}[tp!]
\includegraphics[width=\columnwidth]{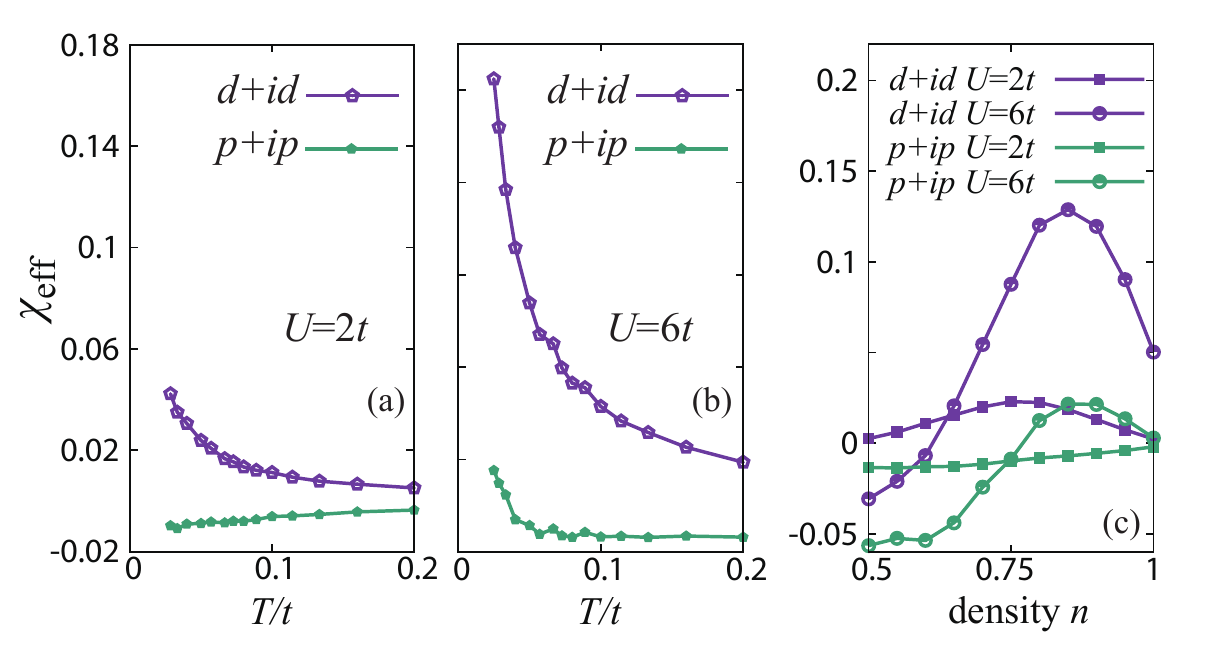}
\caption{(Color online) (a) Effective pairing susceptibilities for $N_c = 3$, $U=2t$ at the VHS density $n=3/4$. (b) Effective pairing susceptibilities for $N_c = 3$, $U=6t$ at the VHS density. (c) Density dependence of the effective pairing susceptibilities at $\beta t = 20$ ($T/t=0.05$). All $d$-wave channels are degenerate, as are the $p$ waves.
The $s$- and $f$-waves are not divergent and hence are not shown here.}
\label{fig:pairingnc3}
\end{figure}

Turning then to the interacting case, we calculated both the temperature and the filling dependence of $\chi_{\text{eff}}$ for different values of $U$ and cluster sizes $N_c$.
In the following, we present explicitly  our results for  $d$- and $p$-wave pairing, which we observe to be the most dominant channels.
In Fig.~\ref{fig:pairingnc3}(a), the temperature dependence of the effective pairing susceptibilities at $U=2t$ and the VHS filling are shown for the $N_c=3$ cluster.  Here, we identify $d+id$ as the dominant pairing channel. However, for $U=6t$ [Fig.~\ref{fig:pairingnc3}(b)], the $p+ip$ triplet channel  increases and also  becomes positive, and furthermore exhibits a tendency to diverge. In order to monitor this behavior as a function of doping, the dependence of $\chi_{\text{eff}}$ on the density $n$ for different interaction strengths at $\beta t = 20$ is shown in Fig.~\ref{fig:pairingnc3}(c). The dome-shaped behavior in the  $d+id$ singlet pairing channel $\chi_{\text{eff}}$  indicates an optimal doping between $n=0.75$ and $0.85$. At $U=6t$, $p+ip$ also exhibits a dome-shaped $\chi_{\text{eff}}$-maximum, even though the amplitude is still lower than for $d+id$.

\begin{figure}[tp!]
\includegraphics[width=\columnwidth]{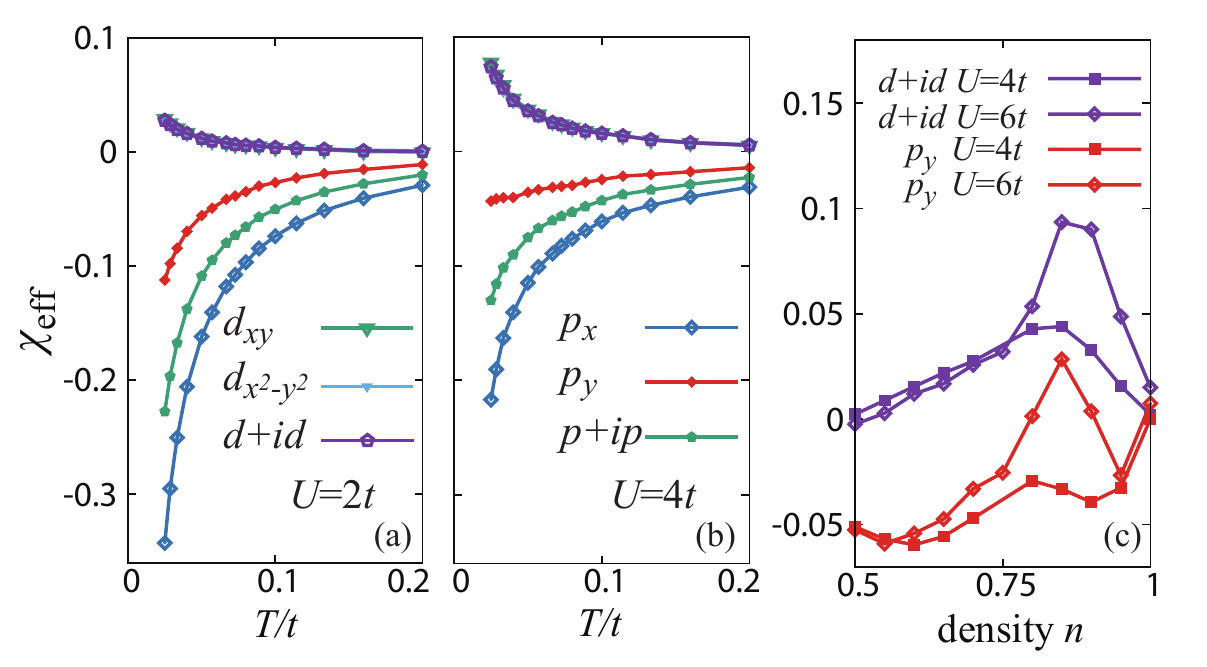}
\caption{(Color online) (a) Effective pairing susceptibilities for $N_c = 4$, $U=2t$ at the VHS density $n=3/4$. (b) Effective pairing susceptibilities for $N_c = 4$, $U=4t$ at the VHS density. Note that the data in panels  (a) and (b) share labels. (c) Density dependence of the effective pairing susceptibility at $\beta t = 20$ ($T/t=0.05$).}
\label{fig:pairingnc4}
\end{figure}

The $N_c=4$ cluster results, shown in Fig.~\ref{fig:pairingnc4}, are to a large extent similar to the $N_c=3$ data: For
weak interactions ($U=2t$ and $4t$), and close to the VHS, the dominant pairing channel is also $d+id$. Upon increasing
the interaction strength to $U=6t$, the $p$-wave effective susceptibility again increases. However, we also find
differences between the results for $N_c=3$ and $N_c=4$. As shown in Fig.~\ref{fig:clusters}(b), the cluster momenta for
$N_c=4$ are the  $\Gamma$ point and the $M$ points. In the non-interacting band structure, these four cluster momenta
are below the Fermi surface for densities $n>0.75$. Thus, the $N_c=4$ cluster does not capture  charge fluctuations
about (below and above) the Fermi surface  for densities $n>0.75$. This is  reflected by the effective pairing
susceptibilities. For example, for  $U=4t$ the effective paring susceptibility at $T/t=0.025$  in the ($d+id$)-wave
channel for $N_c=3$ is $0.154$, while for $N_c=4$, it is $0.074$, i.e., about half  the $N_c=3$ value. Similarly,
due to the deficits of the $N_c=4$ cluster, the $p$-wave channel does not exhibit a tendency to increase for $U=4t$ at the VHS filling [see Fig.~\ref{fig:pairingnc4}(b)].
In Fig.~\ref{fig:pairingnc4}(c), we still observe   a  narrow density regime within which the effective pairing susceptibility for the $p_y$ channel is positive,
but it is smaller than that of the $d+id$ channel. The observed trends suggest that only upon further increasing the
interaction strength might the $p_y$ channel possibly diverge more rapidly than the $d+id$ channel. We note, that the
breaking of the degeneracy among the different $p$-wave channels  is due to finite-size effect; that is, the corresponding form factors have different gap sizes on the finite cluster momenta.

\begin{figure}[tp!]
\includegraphics[width=\columnwidth]{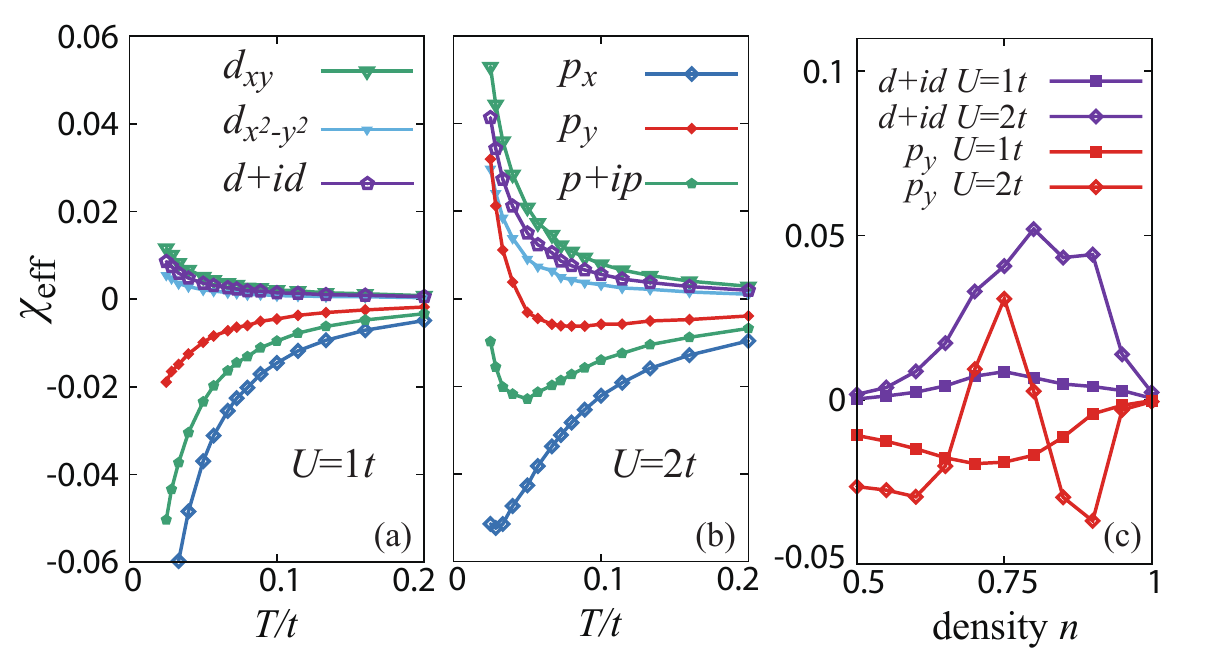}
\caption{(Color online) (a) Effective pairing susceptibility for $N_c = 12$, $U=t$ at the VHS density $n=3/4$. (b) Effective pairing susceptibility for $N_c=12$, $U=2t$ at the VHS density. Note  that the data in  panels (a) and (b) share labels. (c) Density dependence of the effective pairing susceptibility at $\beta t = 40$ ($T/t=0.025$). }
\label{fig:pairingnc12}
\end{figure}

Within the DCA approach, one must study the systematic behavior upon increasing $N_c$ in order to draw  conclusions
about the thermodynamic limit. As demonstrated by comparing the $N_c=3$ and $N_c=4$ results, it is also important to consider clusters that capture the low-energy fluctuations.
We thus also employed  the  $N_c=12$ cluster within the DCA/CT-INT framework, which has cluster momenta that include the
$\Gamma$ point, the two $K$ points, the three $M$ points and six other momenta [ see Fig.~\ref{fig:clusters}(c)].
It  provides a more detailed structure of the pairing symmetry than the $N_c=3$ and $4$ clusters.
Unfortunately, the minus-sign problem  becomes much more severe for $N_c=12$, and we cannot access large values of $U$ for $N_c=12$.
Nevertheless, we can draw interesting observations from the accessible parameter range:
In Fig.~\ref{fig:pairingnc12}(a), for $U=t$, the dominant pairing channel is still $d+id$ near the VHS filling, while in
Fig.~\ref{fig:pairingnc12}(b), we find for $U=2t$ that $p$-wave starts to increase at low temperatures. The $p$-wave channels (in particular $p_y$)  tend to increase more rapidly upon cooling than the $d$-wave channels.
This trend indicates that the $p$-wave  channels compete strongly with  $d$-wave paring at this interaction strength, such that in the intermediate interaction range, there is
an enhanced tendency for $p$-wave triplet pairing to eventually  dominate
over $d$-wave singlet pairing in the thermodynamic limit. To analyze this trend in more detail,
the density dependence of the leading effective pairing susceptibilities is shown in Fig.~\ref{fig:pairingnc12}(c).
The optimal doping range for $d+id$ pairing is consistent with the $N_c=3$ and $4$ results.
In addition,
the enhancement  of the $p$-wave channels upon increasing the interaction strength is quite pronounced on the $N_c=12$ cluster.
Hence, although the minus-sign problem renders us unable to make a definitive statement about whether  $p$-wave triplet pairing will eventually
replace  $d$-wave singlet pairing, the available $N_c=12$ data up to $U=2t$  suggest such a scenario.

\section{Discussions and Conclusions}
Our DCA/CT-INT results are consistent with previous reports that in the weak coupling limit the dominant pairing  is the chiral $d+id$ singlet channel.
However, we find  upon increasing the interaction strength (e.g., for $N_c=12$, $U=2t$) a clear tendency towards a competing $p$-wave triplet pairing.
This finding is  consistent with several recent findings. For example, it has been reported~\cite{Faye_2015_085121},
that in the presence of both on-site interaction and nearest neighbor repulsion, for a wide range of doping around the
VHS, the dominant pairing is a $p$-wave triplet. In Refs.~\onlinecite{Gu_2013_,Gu2014}, where the  infinite-$U$ limit
was considered,  a $p+ip$ wave superconducting ground state was proposed. Our calculation focused on the range of  small and medium strength  interactions,
and indeed suggested the possibility that the dominant pairing channel changes  from $d+id$ to $p$-wave upon increasing the  interaction strength.

For the future, it would be interesting, to allow also for inhomogeneous pairing states within the DCA calculations, in
light of  several recent proposals of superconductivity coexisting with Kekul\'{e} patterns~\cite{Roy_2010,Kunst2015,Faye_2016}. On a more general note, the effect of Hund's  coupling and  spin-orbit coupling could be included in the DCA calculations, given the Hund's coupling induced triplet pairing scenario of
Ref. ~\cite{ Yuan15} as well as  recent NMR experiments on Cu$_x$Bi$_2$Se$_3$, a spin-orbital coupled topological material with moderate electron correlations, which suggest an odd-parity, spin-rotation symmetry breaking triplet pairing state~\cite{Matano2016}.

\section*{Acknowledgments}
The authors thank  H. T. Dang, M. Golor, Z.-C. Gu,  C. Honerkamp,  and T. Ying for helpful discussions. X.Y.X. and
Z.Y.M. are
supported by the Ministry of Science and Technology (MOST) of China under Grant No.  2016YFA0300502, the National
Natural Science Foundation of China (NSFC Grants No.\ 11421092 and No. 11574359), and the National
Thousand-Young-Talents Program of China. X.Y.X. gratefully acknowledges the hospitality of the Institute for Theoretical
Solid State Physics at RWTH Aachen University and support from the Deutsche Forschungsgemeinschaft (DFG) within the
research unit FOR 1807. This work was made possible by generous allocations of CPU time from the Center for Quantum
Simulation Sciences in the Institute of Physics, Chinese Academy of Sciences, and the National Supercomputer Center in
Tianjin. We also acknowledge computing resources at JSC J\"ulich and  RWTH Aachen University with JARA-HPC.

\appendix

\section{Interaction expansion}
\label{app:a}
The partition function for the CT-INT can be obtained as
\begin{eqnarray}
Z & = & \text{Tr}\left[e^{-\beta\hat{H}_{0}}e^{-\int_{0}^{\beta}d\tau\hat{H}_{I}(\tau)}\right]\nonumber \\
 & = & \sum_{k}(-1)^{k}\int_{0}^{\beta}d\tau_{1}\cdots\int_{\tau_{k-1}}^{\beta}d\tau_{k}\text{Tr}\left[e^{-\beta\hat{H}_{0}}\hat{H}_{I}(\tau_{k})\cdots \hat{H}_{I}(\tau_{1})\right]\nonumber \\
 & = & \sum_{\mathcal{C}_{k}}(-\frac{U}{2})^{k}\frac{1}{k!}\prod_{\sigma}\langle\text{T}_{\tau}(\hat{n}_{1\sigma}-\alpha_{1\sigma})(\hat{n}_{2\sigma}-\alpha_{2\sigma})\cdots\nonumber \\
 &  & \cdots(\hat{n}_{k\sigma}-\alpha_{k\sigma})\rangle_{0} \nonumber\\
 & = & \sum_{\mathcal{C}_{k}}(-\frac{U}{2})^{k}\prod_{\sigma}\det\mathbf{D}^{\sigma}(k)
\end{eqnarray}
where we have written the interaction part $\hat{H}_{I}$ of the Hamiltonian Eq.~(1) in the main text as $U/2\sum_{i,s\pm1}\prod_{\sigma}(\hat{n}_{i\sigma}-\alpha_{\sigma}(s))$,
with $\alpha_{\sigma}(s)=1/2+\sigma s(1/2+0^{+})$. The configuration
$\mathcal{C}_{k}=\{[i_{1},\tau_{1},s_{1}]\cdots[i_{k},\tau_{k},s_{k}]\}$.
The $\mathbf{D}^{\sigma}(k)$ matrix has diagonal elements $\mathbf{D}_{pp}^{\sigma}(k)=-\langle T_{\tau}\hat{c}_{i_{p}}^{\dagger}(\tau_{p}^{+})\hat{c}_{i_{p}}(\tau_{p})\rangle_{0}+\alpha_{\sigma}(s_{i})\equiv-g_{0}^{0}(\beta)+\alpha_{\sigma}(s_{i})$
and off-diagnoal elements $\mathbf{D}_{pq}^{\sigma}(k)=-\langle T_{\tau}\hat{c}_{i_{q}}^{\dagger}(\tau_{q}^{+})\hat{c}_{i_{p}}(\tau_{p})\rangle_{0}\equiv g^{0}(p,q)$. The CT-INT solver uses the  cluster excluded Green's function $g^{0}(p,q)$
as input.

\section{DCA loop}
\label{app:b}
The honeycomb lattice has two sites per unit cell, and the DCA self-consistent loop requires more steps for such a
complex lattice than that for simple lattices such as square or triangular lattice. In this appendix, we describe the general DCA scheme we have developed for such more complex lattices.

We define the cluster excluded Green's function in
matrix form $\mathbf{g}^{0}(\vec{R},\tau)$, whose matrix elements are $g_{\alpha\beta}^{0}(\vec{R},\tau)$ with
$\mathbf{R}$ being the distance vector between unit cells,
and $\alpha$, $\beta$ being  indices for the two sublattices, $A$ and $B$.

The DCA loop starts from a non-interacting cluster self-energy $\boldsymbol{\Sigma}_{c}(\vec{K},i\omega_{n})=0$
or the self-energy obtained from second-order perturbation theory. One then uses the cluster self-energy to approximate the lattice self-energy, and the lattice Green's function is
\begin{eqnarray*}
\mathbf{G}^{latt}(\vec{k},i\omega_{n}) & = & \mathbf{G}^{latt}(\vec{K}+\tilde{\vec{k}},i\omega_{n})\\
 & = & \frac{1}{\left(i\omega_{n}+\mu\right)\mathbf{1}-\mathbf{H}_{0}(\vec{K}+\tilde{\vec{k}})-\boldsymbol{\Sigma}_{c}(\vec{K},i\omega_{n})}
\end{eqnarray*}
where
\begin{equation}
\mathbf{H}_{0}(\vec{k})=\left(\begin{array}{cc}
0 & -t\sum_{l=1}^{3}e^{i\vec{k}\cdot \boldsymbol{\delta}_{l}}\\
-t\sum_{l=1}^{3}e^{-i\vec{k}\cdot\boldsymbol{\delta}_{l}} & 0
\end{array}\right)
\end{equation}
and $\boldsymbol{\delta}_{l}$ are the three nearest neighbor vectors, $\boldsymbol{\delta}_{1}=(0,-\frac{1}{\sqrt{3}})$,
$\boldsymbol{\delta}_{2}=(\frac{1}{2},\frac{1}{2\sqrt{3}})$, and $\boldsymbol{\delta}_{3}=(-\frac{1}{2},\frac{1}{2\sqrt{3}})$.
Note that for the honeycomb lattice $\mathbf{G}^{latt}(\vec{k},i\omega_{n})$ and $\boldsymbol{\Sigma}_{c}(\vec{K},i\omega_{n})$ are  $2\times2$ matrices with sublattice indices.
One then needs to prepare the cluster excluded Green's function $\mathbf{g}^{0}(\vec{R},\tau)$ for the CT-INT impurity solver.

In the first step, we coarse grain the lattice Green's function
\begin{equation}
\bar{\mathbf{G}}^{latt}(\vec{k},i\omega_{n})=\frac{1}{N_{\tilde{\vec{k}}}}\sum_{\tilde{\vec{k}}}\mathbf{G}^{latt}(\vec{K}+\tilde{\vec{k}},i\omega_{n})
\end{equation}
with $N_{c}$ being the number of the cluster size (the number of unit cells in a cluster), and $N_{\tilde{\vec{k}}}$
being the number of $\tilde{\mathbf{k}}$ points within each $\vec{K}$ patch.

Then, by using the Dyson equation, the cluster excluded Green's function in $(\mathbf{K},i\omega_n)$ space can be obtained as
\begin{equation}
\mathbf{g}^{0}(\vec{K},i\omega_{n})=\left(\bar{\mathbf{G}}^{latt}(\vec{k},i\omega_{n})^{-1}+\boldsymbol{\Sigma}_{c}(\vec{K},i\omega_{n})\right)^{-1}.
\end{equation}

Finally, $\mathbf{g}^{0}(\vec{K},i\omega_{n})$ needs to be transformed to $\mathbf{g}^{0}(\vec{R},\tau)$ to provide the input for the impurity
solver as
\begin{equation}
\mathbf{g}^{0}(\vec{K},i\omega_{n})\xrightarrow{1}\mathbf{g}^{0}(\vec{K},\tau)\xrightarrow{2}g_{i,j}^{0}(\tau)\xrightarrow{3}\mathbf{g}^{0}(\vec{R},\tau)\label{eq:ftstep}
\end{equation}
For simple lattices, $\mathbf{g}^{0}(\vec{K},i\omega_{n})$ and $\mathbf{g}^{0}(\vec{K},\tau)$ are connected by a Fourier transformation. But for more complex lattices, this requires more steps. We explain steps $1$, $2$ and $3$ in Eq.~(\ref{eq:ftstep}) below.

In step $1$, we perform an infinite Matsubara frequency summation. In order to ensure numerical precision, we  divide the imaginary-time interval into several ranges and consider  them separately. In the following, $n_{c}$ is the frequency cutoff and we take $n_{c}=1000$ in our code. Then,
\begin{equation}
\mathbf{g}^{0}(\vec{K},\tau \in(0,\beta))=-\frac{1}{\beta}\sum_{n=-n_{c}}^{n_{c}-1}\mathbf{g}^{0}(\vec{K},i\omega_{n})e^{-i\omega_{n}\tau},
\end{equation}
and
\begin{eqnarray*}
\mathbf{g}^{0}(\vec{K},\tau=0^{+}) & = & -\frac{1}{\beta}\sum_{n=-n_{c}}^{n_{c}-1}\left(\mathbf{g}^{0}(\vec{K},i\omega_{n})-\frac{1}{i\omega_{n}}\right)e^{-i\omega_{n}\tau}+\\
 &  & - \frac{1}{\beta}\sum_{n=-\infty}^{+\infty}\frac{e^{-i\omega_{n}\tau}}{i\omega_{n}}\\
 & = & -\frac{1}{\beta}\sum_{n=-n_{c}}^{n_{c}-1}\mathbf{g}^{0}(\vec{K},i\omega_{n})+\frac{1}{2},
\end{eqnarray*}
and the periodic boundary condition in the time axis gives
\begin{equation}
\mathbf{g}^{0}(\vec{K},\tau=\beta)=\mathbf{1}-\mathbf{g}^{0}(\vec{K},\tau=0^{+}),
\end{equation}
and
\begin{equation}
\mathbf{g}^{0}(\vec{K},\tau(\in[-\beta,0))=-\mathbf{g}^{0}(\vec{K},\tau+\beta).
\end{equation}

In step $2$, which leads from $\mathbf{g}^{0}(\vec{K},\tau)$ to $g_{i,\alpha;j,\beta}^{0}(\tau)$, we need to perform a modified Fourier transformation
\begin{equation}
g_{i,\alpha;j,\beta}^{0}(\tau)=\frac{1}{N_c}\sum_{\vec{K}}g_{\alpha,\beta}^{0}(\vec{K},\tau)e^{i\vec{K}\cdot\vec{r}_{i}}e^{-i\vec{K}\cdot\vec{r}_{j}},
\end{equation}
using inner cell coordinates in the phase, i.e., $\vec{r}_{i}=\vec{R}_{i}+\mathbf{t}_{\alpha(\beta)}$, where $\vec{R}_{i}$ is the unit cell coordinate and $\mathbf{t}_{\alpha(\beta)}$ is the inner cell coordinate: $\mathbf{t}_{\alpha}=(0,0)$ and $\mathbf{t}_{\beta}=(0,1/\sqrt{3})$, for the two sublattices. 

In step $3$, leading from $g_{i,\alpha;j,\beta}^{0}(\tau)$ to $g_{\alpha,\beta}^{0}(\vec{R},\tau)$, we make use of  translational symmetry and perform a constrained summation
\begin{equation}
g_{\alpha\beta}^{0}(\vec{R},\tau)=\frac{1}{N_{c}}\sum_{\substack{\vec{R}_{i}-\vec{R}_{j}=\vec{R}}
}g_{i,\alpha;j,\beta}^{0}(\tau).
\end{equation}

\section{Two particle Green's function}
\label{app:c}
To calculate correlation functions, we need to evaluate the two-particle Green's functions within the CT-INT, $\langle G_{\alpha_{1}\beta_{1}}^{\sigma}(P_{1},P_{1}')G_{\alpha_{2}\beta_{2}}^{\sigma}(P_{2,}P_{2}')\rangle$,
where $G_{\alpha\beta}(P,P')$ is defined as
\begin{equation}
\begin{split}
   &\   G_{\alpha\beta}\left(P(\vec{K},i\omega_{n}),P'(\vec{K}',i\omega_{n}')\right) \\
=  &\   \mathbf{g}_{\alpha,\beta}^{0}(\vec{K},i\omega_{n})\delta_{\vec{K},\vec{K}'}\delta_{i\omega_{n},i\omega_{n}'} \\
   &  -\mathbf{g}_{\alpha,\gamma}^{0}(\vec{K},i\omega_{n})\Gamma_{\gamma,\eta}(\vec{K},i\omega_{n};\vec{K}',i\omega_{n}')\mathbf{g}_{\eta,\beta}^{0}(\vec{K}',i\omega_{n}')
\end{split}
\end{equation}
with
\begin{equation}
\begin{split}
  & \Gamma_{\alpha,\beta}(\vec{K},i\omega_{n};\vec{K}',i\omega_{n}') \\
= & -\frac{T}{N_{c}}\sum_{\substack{i,j}}e^{-i\vec{K}\cdot\vec{r}_{i}}e^{i\omega_{n}\tau_{i}}\mathbf{M}(k)_{i,\alpha;j,\beta}e^{-i\omega_{n}'\tau_{j}}e^{i\vec{K}'\cdot\vec{r}_{j}}
\end{split}
\end{equation}
where $\mathbf{M}(k)=\mathbf{D}(k)^{-1}$ and $\vec{r}_{i}=\vec{R}_{i}+\mathbf{t}_{\alpha(\beta)}$, where $\vec{R}_{i}$ is the unit cell coordinate and $\mathbf{t}_{\alpha(\beta)}$ is the inner cell coordinate.

\section{Details on calculating the pairing susceptibility}
\label{app:d}
In the DCA formalism, we need to distinguish the cluster pairing susceptibility and the lattice pairing susceptibility.
The real physical quantities, the lattice susceptibilities $\bar{\chi}$, are obtained with
\begin{equation}
\bar{\chi}=\frac{\bar{\chi}^{0}}{1-\Gamma'\bar{\chi}^{0}}\label{eq:lattice_chi}
\end{equation}
where $\bar{\chi}^{0}$ is the coarse-grained non-interacting susceptibility,
and $\Gamma'$  the irreducible vertex.  Within the DCA approximation,  the irreducible vertex $\Gamma'$ in the lattice susceptibility and the $\Gamma_{c}$ in the cluster susceptibility are equivalent once they are coarse-grained to the cluster level, so that
\begin{equation}
\left(\bar{\chi}^{0}\right)^{-1}-\left(\bar{\chi}\right)^{-1}=\Gamma'=\Gamma_{c}=\left(\chi_{c}^{0}\right)^{-1}-\left(\chi_{c}\right)^{-1}.\label{eq:bse}
\end{equation}
The cluster pairing susceptibility matrix $\chi_{c}\left(P,P',Q=0\right)$
is defined as
\begin{equation}
\left(\begin{array}{cc}
\langle G_{11}^{\uparrow}(-P,-P')G_{22}^{\downarrow}(P,P')\rangle & \langle G_{12}^{\uparrow}(-P,-P')G_{21}^{\downarrow}(P,P')\rangle\\
\langle G_{21}^{\uparrow}(-P,-P')G_{12}^{\downarrow}(P,P')\rangle & \langle G_{22}^{\uparrow}(-P,-P')G_{11}^{\downarrow}(P,P')\rangle
\end{array}\right)
\end{equation}
and the non-interacting pairing susceptibility matrix $\chi_{c}^{0}(P,P',Q=0)$
is defined as
\begin{equation}
\left(\begin{array}{cc}
G_{11}^{\uparrow}(-P)G_{22}^{\downarrow}(P) & G_{12}^{\uparrow}(-P)G_{21}^{\downarrow}(P)\\
G_{21}^{\uparrow}(-P)G_{12}^{\downarrow}(P) & G_{22}^{\uparrow}(-P)G_{11}^{\downarrow}(P)
\end{array}\right)\delta_{P,P'}.
\end{equation}
In the above equations, $1$ and $2$ denote sublattice indices and $P=(\mathbf{K},i\omega)$, $P'=(\mathbf{K'},i\omega')$ and
$Q=(\mathbf{Q},i\nu)$ are four-vectors of cluster momentum and Matsubara frequency. Based on $\chi_{c}$ and $\chi_{c}^{0}$, by using Eq. (\ref{eq:bse}),
we  then get the irreducible vertex $\Gamma_{c}$, equivalently, $\Gamma'$.
Then using Eq. (\ref{eq:lattice_chi}), we obtain the lattice pairing susceptibility
$\bar{\chi}$. Finally, we get the effective pairing susceptibility from
\begin{eqnarray*}
\chi_{eff}^{\eta}(T) & = & \frac{1}{\beta}\sum_{P,P',Q=0}\langle\bar{\Phi}_{\eta}(\vec{K})|\bar{\chi}(P,P',Q=0)-\\
 &  & \bar{\chi}^{0}(P,P',Q=0)|\bar{\Phi}_{\eta}(\vec{K}')\rangle/\sum_{\vec{K}}\langle\bar{\Phi}_{\eta}(\vec{K})|\bar{\Phi}_{\eta}(\vec{K})\rangle,
\end{eqnarray*}
where $\bar{\Phi}_{\eta}(\vec{K})$ and $\bar{\Phi}_{\eta}(\vec{K}')$ are coarse-grained
form factors as shown in the Fig. 1(d) in the main text.

\section{Relation to the pairing vertex eigenvalue analysis}
\label{app:e}
An eigenvalue analysis of the pairing vertex secular equation requires us to solve  the eigenvalue problem
\begin{equation}
\Gamma \chi ^0 | \phi_i \rangle  = \lambda_i  |   \phi_i \rangle.
\end{equation}
In terms of the pairing vertex,
the effective pairing susceptibility is given as
\begin{equation}
\begin{split}
\chi_{eff} &  =    \chi - \chi ^0   \\
 &  =   \chi ^0 \left(  1 - \Gamma \chi ^0 \right) ^{-1} -\chi ^0.
\end{split}
\end{equation}
Inserting the diagnoal representation of $\Gamma \chi^0$ into the above equation, one obtains
\begin{equation}
\chi_{eff} = \chi^0 \sum_i  \frac {\lambda_i} { 1-\lambda_i}  |\phi_i \rangle \langle \phi_i |.
\label{eq:chieff}
\end{equation}
The pairing susceptibility is similarly found to be given as
\begin{equation}
\chi = \chi^0 \sum_i \frac {1} { 1-\lambda_i}  |\phi_i \rangle \langle \phi_i |.
\end{equation}
When the largest eigenvalue of the pairing vertex secular equation approaches 1, both the pairing susceptibility and the
effective pairing susceptibility diverge. Usually, a positive effective pairing susceptibility indicates an enhancement
of the pairing correlations due to the interaction vertex~\cite{White_1989}. Comparing the effective pairing
susceptibilities is thus equivalent to comparing the strength of the vertex corrections, and allows one to identify the dominant pairing channel.

\bibliography{main}

\end{document}